\documentclass{article}
\def\M{{\cal M}}
\def\C{{\cal C}}
\def\E{{\cal E}}
\def\fslash{{\kern 2pt\raise 1.5pt \hbox{$\backslash$} \kern -7pt}}
\def\boxit#1{\vbox{\hrule\hbox{\vrule\kern3pt\vbox{\kern3pt#1\kern3pt}\kern3pt\vrule}\hrule}}
\font\dias=cmr10
\begin{document}
\markboth{Brian P. Dolan}{Non-commutative Complex Projective Spaces
and the Standard Model}

\title{\hfill\raise 20pt\hbox{\dias DIAS-STP-03-06}\\
Non-commutative Complex Projective Spaces\\ 
And The Standard Model\footnote{Talk presented at the conference
\lq\lq Space-time and Fundamental Interactions: Quantum Aspects''
in honour of A.P.~Balachandran's 65th birthday, Vietri sul Mare, Salerno, Italy
26th-31st May, 2003}}
\author {Brian P. Dolan\\
Department of Mathematical Physics\\ National University of Ireland\\ 
Maynooth, Co. Kildare, Ireland\\
and \\
Dublin Institute for Advanced Studies\\
10, Burlington Rd., Dublin, Ireland\\
{\tt bdolan@thphys.may.ie}\\}

\maketitle

\begin{abstract}

The standard model fermion spectrum, including
a right handed neutrino, can be obtained as a zero-mode
of the Dirac operator on a space which is the product of
complex projective spaces of complex dimension two and three.
The construction requires the introduction of topologically 
non-trivial background gauge fields.  By borrowing from ideas in
Connes' non-commutative geometry and making the complex spaces
\lq fuzzy' a matrix approximation to the fuzzy space allows
for three generations to emerge.  The generations
are associated with three copies of space-time.  Higgs'
fields and Yukawa couplings can be accommodated in the usual
way.

\end{abstract}

\section{Introduction}	
Current descriptions of the force of gravity and the fundamental
interactions of particle physics are set in the language of
differential geometry and fibre bundles.  A unified description
of gravity and gauge theories has long been one of the main
goals of modern theoretical physics and superstring theory is
currently the most popular framework for this endeavour.
Nevertheless it may be worthwhile pursuing other avenues of investigation
and a suggestion is made here, based on an observation
of a curious connection between the standard model of particle
physics and the Atiyah-Singer index theorem on complex
projective spaces.

The motivation is to bring the geometrical descriptions
of general relativity and Yang-Mills theory closer together.
In a geometrical approach to fundamental interactions
physical fields  are
tensors associated with the tangent space of an underlying
four dimensional manifold, ${\M}$:
\begin{table}[ht]
{\begin{tabular}{@{}ll@{}}
{\bf General Relativity}  &   {\bf Yang-Mills Theory}  \\
Manifold: $\M$ & Manifold: $\M$ \\
Tangent Bundle: $T\M$  & Vector Bundle: $\E(M)$\\
Lorentzian Metric: $g$   & Connection: $A$ \\
Spin Connection: $\omega$  & Curvature: $F$  \\
Curvature: $R$ & Local Gauge group:\\
Local tangent space rotations: & e.g. $U(n)$ or $SU(3)\times SU(2)\times U(1)$ \\
$SO(3,1)$ (or $Sl(2,{\bf C})$)   
\end{tabular}}
\end{table}

In Kaluza-Klein theories one brings these structures together by
taking a compact co-set space, $\C\cong G/H$, of small radius
and extending $\M$ to $\M\times \C$.  The gauge group is identified
with the isometry group $G$ of $\C$.  In the spirit of general
relativity it is perhaps more natural to identify the holonomy
group $H$ of $\C$ with the Yang-Mills gauge group, since it is the 
holonomy group of $\M$ that takes the centre stage in gravity.
If $\C$ is a complex manifold, with $n$ complex dimensions
and an hermitian metric, it has holonomy group $U(n)$ in general
and so might furnish a $U(n)$ gauge theory.  

If the Dirac operator on $\M\times\C$ decomposes as a direct sum
\begin{equation}
i\fslash D = i\fslash D_\M \otimes {\bf 1}+\gamma_5\otimes i\fslash D_\C,
\label{Dirac}
\end{equation}
where $\gamma_5=\gamma^0\gamma^1\gamma^2\gamma^3$ is the chirality operator
on $\M$, then eigenspinors $\zeta$ of the internal Dirac operator,
\begin{equation}
i\fslash D_\C\zeta=\lambda\zeta,
\end{equation}
will have eigenvalues of the order of the Ricci scalar on $\C$.
In the spirit of Kaluza-Klein theory, if $\C$ has
a very large curvature, only the zero eigenstates
will contribute to the low energy spectrum seen in $\M$.
It was shown by Witten \cite{Witten} that the chiral fermion
spectrum of the standard model in $\M$ cannot be obtained in this
way purely from the metric and spin connection on $\C$ and this
essentially killed the Kaluza-Klein programme in the early '80\,s.
Some extra ingredient is needed and we shall avoid Witten's theorem
by introducing fundamental gauge fields.

We are thus led to an investigation of the zero modes of the Dirac
operator on complex manifolds in the presence of background gauge
fields but
we first discuss Clifford algebras and
complex vector spaces.

\section{Clifford Algebras on Complex Vector Spaces}
Let $z^\mu$, $\mu=1,\ldots,n$ be complex co-ordinates on a
complex manifold with complex dimension $n$.  Dirac fermions
have $2^n$ components and Weyl fermions $2^{n-1}$ components.
The $\gamma$-matrices can be chosen so that
\begin{equation}
\{\gamma^a,\gamma^b\}=\{\gamma^{\bar a},\gamma^{\bar b}\}=0,\qquad
 \{\gamma^a,\gamma^{\bar b}\}=2\delta^{a\bar b}
\end{equation}
with $a,b$ indices labelling an orthonormal basis.
This Clifford algebra is isomorphic to the algebra of 
of $n$-fermionic creation and annihilation operators \cite{GSW}.
Because of the fermionic nature of the creation and
annihilation operators, $b^a={1\over\sqrt 2}\gamma^a$ and  
$\left(b^a\right)^\dagger={1\over\sqrt 2}\gamma^{\bar a}$,
the Fock space is $2^n$-dimensional.
Denoting the vacuum state by $|\Omega>$ 
we can construct a basis for the Fock space
\begin{equation}
|\Omega>,\qquad |\Omega^{\bar a}>=\gamma^{\bar a},\qquad
|\Omega^{\bar a\bar b}>=\gamma^{\bar a\bar b}=-|\Omega^{\bar b\bar a}>,
\qquad |\Omega^{\bar a\bar b\bar c}>=\gamma^{\bar a\bar b\bar c},\quad 
\hbox{etc.,}
\end{equation}
where $\gamma^{\bar a\bar b}=(1/2)[\gamma^{\bar a},\gamma^{\bar b}]$,
and higher rank products are similarly anti-symmetrised.
A Dirac spinor can then be expanded as
\begin{equation}
\zeta=\phi|\Omega> +\phi_{\bar a}|\Omega^{\bar a}>
+\phi_{\bar a\bar b}|\Omega^{\bar a\bar b}> +\cdots +
\phi_{\bar a_1\cdots\bar a_n}
|\Omega^{\bar a_1\cdots \bar a_n}>.
\end{equation}
This can be decomposed into two Weyl spinors
\begin{eqnarray}
\zeta_+ &=&\phi|\Omega> 
+\ \phi_{\bar a\bar b}|\Omega^{\bar a\bar b}> +\cdots\\
\zeta_-&=&\phi_{\bar a}|\Omega^{\bar a}>+\ \phi_{\bar a\bar b\bar c}
|\Omega^{\bar a\bar b\bar c}>+\cdots.
\end{eqnarray}
We can thus read off how the spinor components $\phi_{\bar{a}_1\cdots \bar{a}_i}$
transform under the $SU(n)$ part of the holonomy:
\begin{eqnarray}
\phi \qquad & \qquad {\bf 1} & \qquad SU(n) \quad \hbox{\footnotesize SINGLET}\nonumber\\
\phi_{\bar a} \qquad & \qquad \overline {\bf n} & \qquad \hbox{\footnotesize ANTI-FUNDAMENTAL}\nonumber\\
\vdots \qquad && \\
\phi_{\bar{a}_1\cdots \bar{a}_{n-1}} & \qquad {\bf n} & \qquad \hbox{\footnotesize FUNDAMENTAL}\nonumber\\
\phi_{\bar{a}_1\cdots \bar{a}_{n}} \ & \qquad {\bf 1}&\qquad \hbox{\footnotesize SINGLET.}\nonumber
\end{eqnarray}
The $U(1)$ charges are more subtle, since the creation and annihilation
operators have $U(1)$ phases.  Normalise the charge so that $b^a$
has charge $+1$ --- then the vacuum can have a charge, which will
be denoted by $k$ for the moment, and we shall fix later.  The $U(1)$
charges of our Fock space basis are now
\begin{equation}
|\Omega> \sim k; \qquad |\Omega^{\bar a}> \sim k-1; \qquad
\cdots\qquad
|\Omega^{{\bar a}_1\cdots {\bar a}_n}> \sim k-n.
\end{equation}

As an example consider the case of a 4-dimensional space with $n=2$.
Without any complex structure the holonomy group is $SO(4)\approx
SU(2)\times SU(2)$ but this can be restricted to $SU(2)\times U(1)$
when a complex structure and a compatible hermitian metric are introduced.
From the decomposition 
under $SU(2)\rightarrow U(1)$ 
\begin{equation}
{\bf 2}\rightarrow {\bf 1}_1 +{\bf 1}_{-1}
\end{equation}
we see the following structure:
\begin{equation}
\begin{array}{ccccc}
\underline{SO(4)\approx SU(2)\times SU(2)} &\ \longrightarrow\ & 
\underline{SU(2)\times U(1)} &\ \longrightarrow\ & 
\underline{SU(2)\times U(1)}\\
&&&&\\
({\bf 2},{\bf 1})+({\bf 1},{\bf 2}) & \longrightarrow & 
{\bf 2}_0 + ({\bf 1}_1+{\bf 1}_{-1}) &
\longrightarrow & {\bf 2}_{-1}+ ({\bf 1}_0+{\bf 1}_{-2})\\
\hbox{\footnotesize DIRAC} && \ |\Omega^{\bar a}> +
(|\Omega>+ |\Omega^{\bar a\bar b}>) && 
\kern -20pt\left(\matrix{\nu_L \cr e_L \cr}\right) \quad \nu_R \quad e_R \\
&&(k=1)&&(k=0)\\
\end{array}
\label{EW}
\end{equation}
It is natural
to take $k=1$, as indicated in the middle column.  But if we
shift $k$ to zero we see that the states in the last column have
the quantum numbers of the electro-weak sector of the standard model.
I first became aware of this assignment of quantum numbers to
the components of a spinor for $n=2$ from Balachandran \cite{fuzzyCP2}.

Now consider a 6-dimensional space with $n=3$.
Without any complex structure the holonomy group is $SO(6)\approx
SU(4)$ but this can be restricted to $SU(3)\times U(1)$
when a complex structure and a compatible hermitian metric are introduced:
\begin{equation}
\begin{array}{ccc}
\underline{SO(6)\approx SU(4)} &\qquad\longrightarrow\qquad & \underline{SU(3)\times U(1)}\\
&&\\
{\bf 4}+\overline {\bf 4} & \longrightarrow &
\left({\bf 3}_{-1/2} + {\bf 1}_{3/2}\right) +
\left(\overline{\bf 3}_{1/2} + {\bf 1}_{-3/2}\right)  \\
&&\\
\hbox{\footnotesize DIRAC} && (|\Omega^{\bar a\bar b}>\ |\Omega> )
\quad(|\Omega^{\bar a}>|\Omega^{\bar a\bar b\bar c}>) \\ 
&&(k=3/2)\\
\end{array}
\end{equation}

In this instance we see that $k=3/2$ gives the correct
$U(1)$ charges for the decomposition 
${\bf 4} \rightarrow {\bf 3}_{-1/2}  +{\bf 1}_{3/2}$ (the normalisation of
the $U(1)$ charge is at our disposal).
If $k$ is shifted to zero a single Weyl fermion reduces to
\begin{equation}
{\bf 4} \rightarrow {\bf 3}_{-2}  +{\bf 1}_0.
\label{QCD}
\end{equation}

If we now take the tensor product of
a Dirac fermion for $n=2$ with $k=0$, dividing the
$U(1)$ charges by two to give 
${\bf 2}_{-1/2}+ ({\bf 1}_0+{\bf 1}_{-1})$, and and Weyl
fermion for $n=3$, dividing the $U(1)$ charges by $-3$
to give ${\bf 3}_{2/3}  +{\bf 1}_0$, we find
\begin{eqnarray}
\left({\bf 3}_{2/3} \right. &+&\left.{\bf 1}_0\right)^+\otimes
\left(({\bf 2}_{-1/2})^-+ ({\bf 1}_0+{\bf 1}_{-1})\right)^+
=\nonumber\\
&&\kern -20pt ({\bf 3},{\bf 2})^-_{1/6}+({\bf 3},{\bf 1})^+_{2/3}
+({\bf 3},{\bf 1})^+_{-1/3}+({\bf 1},{\bf 2})^-_{-1/2}
+({\bf 1},{\bf 1})^+_{0}+({\bf 1},{\bf 1})^+_{-1},
\label{SM}
\end{eqnarray}
with the superscript $\pm$ denoting the chirality.
The fermion spectrum of the standard model emerges, including 
a right-handed neutrino.
This structure can be summarised by putting the 16 fermion states into
a $4\times 4$ matrix,
\begin{equation}
\zeta=\left(\begin{array}{llll}
u_L^r & u_L^g & u_L^b & \nu_L \\
u_d^r & d_L^g & d_L^b & e_L \\
u_R^r & u_R^g & u_R^b & \nu_R \\
d_R^r & d_R^g & d_R^b & e_R \\
\end{array}\right),
\label{square}
\end{equation}
and the action of $SU(3)\times SU(2)\times U(1)$ on this matrix
is represented by
\begin{equation}
\zeta\rightarrow g_2 \zeta g_3^*
\label{twothreestar}
\end{equation}
where $g_2\in U(2)$ and $g_3\in U(3)$ --- the $U(1)$ action is just
the difference of the $U(1)$'s in $U(3)$ and $U(2)$, if the $k=0$ charge
assignments in (\ref{EW}) and (\ref{QCD}) are multiplied
by ${1\over 2}$ and ${1\over 3}$ respectively and then subtracted as
implied by (\ref{twothreestar}).

\section{Global Spinor Fields}

The considerations of the last section were purely algebraic
and it is a much more involved question to decide whether these structures
can be defined globally on a given complex space.  Indeed it is well
known that the complex projective space $CP^2$ ($n=2$) does not
admit a globally well defined spin structure \cite{LawsonMichelson}.
Fortunately 
even when a complex manifold manifold does not admit a spin structure
it always admits a spin$^c$ structure, obtained by introducing
a topologically non-trivial $U(1)$ background gauge field.
An essential tool in determining the zero modes of the Dirac
operator in a background field is the Atiyah-Singer index theorem
and an analysis of the index on $CP^n$ was given in Ref.~\cite{CNBD} 
(the particular case of a $U(1)$ field on $CP^2$ was first
analysed, in physicists language, in Ref.~\cite{HawkingPope}).

 Here we quote the results in Ref.~\cite{CNBD} for the index of the
Dirac operator for a fermion on $CP^n$ in a background $U(n)$
gauge field obtained by identifying the gauge connection with the
spin connection, $A\approx\omega$. For a fermion which is 
a $SU(n)$ singlet and has $U(1)$ charge $Y_{(n)}=q$ the index is
\begin{equation}
\nu_q={(q+1)\cdots (q+n)\over n!},
\label{singletindex}
\end{equation}
where the $U(1)$ charge is normalised so that the fundamental unit of charge
is an integer and $q=0$ is the spin$^c$ structure.\footnote{Since the Euler
characteristic of $CP^n$ is $\chi_{n}=n+1$ the first Chern class of
the spin connection is $n+1$, by the Gauss-Bonnet theorem.  Thus
a \lq natural' unit of charge might be $1/\chi_n=1/(n+1)$ ---
in these units a fermion of unit charge would couple to the gauge connection
with the same weight as to the spin connection. 
Relative to this \lq natural' unit the charges $q$ in the text have 
been scaled by $\chi_n$
so as to make the fundamental unit of charge equal
to unity rather than $1/\chi_n$.}

A fermion in the fundamental representation, ${\bf n}$, of $SU(n)$,
with $U(1)$ charge $Y_{(n)}=q+{1\over n}$, has index
\begin{equation}
\nu_{q,{\bf n}}={(q+1)\cdots (q+n-1)(q+n+1)\over (n-1)!}
\end{equation}
(the $U(1)$ charge is $q+{1\over n}$ because a $U(n)$ instanton
on $CP^n$ has first Chern class one, so the $U(1)$
generator is ${1\over n}{\bf 1}$ where ${\bf 1}$ is the unit $n\times n$
matrix, Ref.~\cite{CNBD}).

Using these formulae it can be shown that the algebraic structure
described in section 2 is, in fact, global on $CP^2\times CP^3$ \cite{CNBD}.
We construct a Dirac spinor on $CP^2$ and positive chirality Weyl
spinor on $CP^3$ by taking the following combinations:
\begin{eqnarray}
  \bullet\quad & CP^2 &\ - \ 
    \left\{
    \begin{array}{lllll}
      SU(2)& 
      \hbox{\footnotesize SINGLET WITH} & q=0, & \nu_0=1, & Y_{(2)}=0 \\
      SU(2)& 
      \hbox{\footnotesize SINGLET WITH} & q=-3,\ & \nu_{-3}=1,& Y_{(2)}=-3 \\
      SU(2)& 
      \hbox{\footnotesize DOUBLET WITH\ } & q=-2,\ & \nu_{-2,{\bf 2}}=-1,\ & 
      Y_{(2)}=-{3\over 2} \\
    \end{array}\right. \nonumber\\
  \bullet\quad & CP^3 &\ - \ 
    \left\{
    \begin{array}{lllll}
      SU(3)& 
      \hbox{\footnotesize SINGLET WITH} & q=0, & \nu_0=1, & Y_{(3)}=0 \\
      SU(3)& 
      \hbox{\footnotesize TRIPLET WITH}\ \ \; & q=-3,\ \ \;& \nu_{-3,{\bf 3}}=1,\  
                & Y_{(3)}=-{8\over 3}. \\
\end{array}\right.\nonumber\\
\end{eqnarray}
Defining the hypercharge as 
\begin{equation}
Y:={1\over 3}Y_{(2)}-{1\over 4}Y_{(3)}
\label{Ycharge}
\end{equation}
the tensor product of these zero-modes is
\begin{eqnarray}
\left({\bf 3}_{2/3} \right. &+&\left.{\bf 1}_0\right)^+\otimes
\left(({\bf 2}_{-1/2})^-+ ({\bf 1}_0+{\bf 1}_{-1})\right)^+
=\nonumber\\
&&\kern -20pt ({\bf 3},{\bf 2})^-_{1/6}+({\bf 3},{\bf 1})^+_{2/3}
+({\bf 3},{\bf 1})^+_{-1/3}+({\bf 1},{\bf 2})^-_{-1/2}
+({\bf 1},{\bf 1})^+_{0}+({\bf 1},{\bf 2})^+_{-1},\nonumber\\
\end{eqnarray}
which is precisely that of (\ref{SM}).

The structure here was obtained by identifying the spin connection
with the gauge connection, $A\approx \omega $, but the existence 
of zero-modes
does not require this, the connections can be varied independently and
the zero-modes are guaranteed to persist by the index theorem --- provided
the connections are kept within their topological classes.
Unlike standard Kaluza-Klein theory the structure here does
not rely on any special isometry symmetry being present.

\section{Harmonic Expansion of zero-modes}

The zero-modes of the Dirac operator on $CP^n$ are closely
related to the representation theory of $SU(n+1)$.  Consider,
for example, the index for fermions on $CP^2$ 
which are $SU(2)$ singlets with $U(1)$
charge $q$. Equation (\ref{singletindex}), with $n=2$,
gives
\begin{equation}
\nu_q={(q+1)(q+2)\over 2}.
\end{equation}
For $q=0,1,2,3,\ldots$ this gives $\nu_q=1,3,6,10,\ldots$ and it is no
coincidence that these are the dimensions of irreducible 
representations of $SU(3)$ --- more specifically the symmetric 
irreducible representations.  The same is true for $CP^3$.  The 
representations required in section 3 all
had index with $|\nu|=1$ and so correspond to singlets of $SU(3)$
on $CP^2$ and singlets of $SU(4)$ on $CP^3$.  This has the the
immediate consequence that there exists a metric and a gauge connection in the
relevant topological sector for which these zero-modes are {\it constant}
spinors.  Although the existence of zero-modes, being a topological statement,
does not require any specific metric on $CP^n$, 
one can choose to work with the $SU(n+1)$ symmetric metric
(the Fubini-Study metric \cite{EGH}).  There is then a linear
combination of spinor components for which the gauge connection
exactly cancels the spin connection in the Dirac equation \cite{ASQHE}
\begin{equation}
i\fslash D_\C\zeta =i\gamma^a{e^\mu}_a\left(\partial_\mu +\omega_\mu +
i\bigl(\hbox{$\frac{p}{\chi_n}$}\bigr)A_\mu\right)\zeta,
\label{DiracC}
\end{equation}
so a solution is to take $\zeta=const$ ($p$ here is the $U(1)$
charge coupling to the $U(1)$ gauge connection --- there is a contribution
to $q$ from the spin connection too \cite{CNBD}).

The fact that the zero-modes have such a simple structure suggests
introducing a new ingredient to the construction presented here.
The full space $\M\times \C$ discussed in the introduction is
a fourteen dimensional space and so one cannot hope that the standard
model would be renormalisable in this space, but we can expect
a renormalisable theory if the internal space $\C$ had only a finite
number of degrees of freedom rather than the infinite number of
a continuum manifold.  Borrowing from ideas of Connes \cite{Connes} 
this suggests that one might replace the complex projective
spaces by finite matrix approximations --- fuzzy $CP^n$'s \cite{CPN}.
We are thus led to a picture of the standard model involving non-commutative
geometry similar in spirit, but different in detail, to the
Connes-Lott model \cite{ConnesLott}.
Fuzzy $CP^2$ has finite matrix approximations of dimension
$1 \times 1$, $3\times 3$, $6\times 6, \ldots, d_L\times d_L$, where
$d_L=(L+1)(L+2)/2$ 
are the dimensions of the symmetric representations of $SU(3)$,
and fuzzy $CP^3$ similarly requires matrix algebras whose size is
dictated by the symmetric representations of $SU(4)$ \cite{CPN}.
If we replace the continuum $CP^2\times CP^3$ with its fuzzy
version, $CP_F^2\times CP_F^3$, then chiral spinors become matrices
\begin{equation}
\zeta(z)\qquad \rightarrow \qquad \hat\zeta\in{Mat}_{d_L}\otimes{\bf C}^{2(n-1)},
\label{fuzzyspinor}
\end{equation}
where ${Mat}_{d_L}$ is the algebra of $d_L\times d_L$ matrices,
$z$ is a point in $CP^n$ and ${\bf C}^{2(n-1)}$ is chiral spin space.
In fact, for the constant spinors that we require for our zero-modes,
we only need the trivial representations, $d_L=1$, for both $CP^2_F$
and $CP^3_F$.

\section{Generations and Yukawa Couplings}

The construction described in the previous section only
provides one generation of the standard model, because the
index of the Dirac operator is $\pm 1$.  An obvious question
is whether or not it is possible to obtain 3 generations in some way.
Since the inclusion of generations requires an $SU(3)$ symmetry
it is natural to ask whether or not the $SU(3)$ isometry
group of  $CP^2\cong SU(3)/U(2)$ might be able to provide the
extra generations, so let us focus on $CP^2_F$ (fuzzy $CP^2$
was analysed in detail in Ref.~\cite{fuzzyCP2}).
Since the $SU(3)$ generation symmetry
is broken in the real world, let us assume that it is
broken in our model too.  For example we could deform
the Fubini-Study metric so that it no longer has $SU(3)$
as its group of Killing vectors.  This will change the spin
connection $\omega_\mu$ in equation ({\ref{DiracC}), while leaving the gauge
connection $A_\mu$ unchanged.  We no longer have exact cancellation
between the gauge and spin connections, there are still zero-modes
but they must become non-trivial functions of position, $\partial_\mu \zeta\ne 0$.  An immediate consequence of this is that singlets alone are
no longer sufficient for a harmonic expansion of $\zeta(z)$, higher
dimensional representation of $SU(3)$ must be included.  
The simplest possibility is that $d_L$ in (\ref{fuzzyspinor}) is three
and $\hat\zeta$ is then a $3\times 3$ matrix,
with $SU(3)$ representation content 
$\bar{\bf 3}\times {\bf 3}= {\bf 1}+{\bf 8}$, whose entries depend
on the parameters describing the metric deformation.
The Dirac operator (\ref{Dirac}) on $\M\times\C$ acts on
spinors $\Psi(x)=\psi(x)\otimes\hat\zeta$ where $\psi(x)$ is spinor on $\M$,
($x\in\M$), and we are led to consider
\begin{equation}
\overline{\Psi}i\fslash D\Psi(x) = 
\left(\overline {\psi(x)}\otimes \hat\zeta^\dagger\right)
\Bigl(i\fslash D_\M\otimes {\bf 1}\Bigr) \left(\psi(x)\otimes \hat\zeta\right)
=\left(\overline {\psi(x)}i\fslash D_\M \psi(x)\right)\otimes \hat\zeta^\dagger\hat\zeta.
\end{equation}
Now $\hat\zeta^\dagger \hat\zeta$ is a $3\times 3$ hermitian matrix
and so can be diagonalised by an $SU(3)$ transformation and 
the three eigenstates would look like three generations in $\M$.

One can introduce Yukawa couplings in the usual way: $\hat \zeta$ represents
the set of zero-modes which we shall denote by
\begin{equation}
	\hat Q_L:=\left(
	\begin{array}{c}
		\hat U_L \\
		\hat D_L \\
	\end{array}
	\right)_{1/6}\qquad
	\begin{array}{c}
		(\hat U_R)_{2/3}\\  
		(\hat D_R)_{-1/3}\\  
	\end{array}
	\qquad
	\hat L_L:=\left(
	\begin{array}{c}
		\hat N_L \\
		\hat E_L \\
	\end{array}
	\right)_{-1/2}\qquad
	\begin{array}{c}
		(\hat E_R)_{-1}\\
		(\hat N_R)_0,\\
	\end{array}
\end{equation}
where each of these states is a $3\times 3$ matrix (the subscript
denotes the hypercharge).
Introduce Higgs scalars
\begin{equation}
\Phi=\left(\matrix{\varphi_+ \cr \varphi_0 \cr}\right)_{1/2}
\qquad
\Phi_C=(i\sigma_2)\Phi^*=
\left(\matrix{(\varphi_0)^* \cr -\varphi_- \cr}\right)_{-1/2},
\end{equation}
which transform under $SU(2)\times U(1)$ as ${\bf 2}_{1/2}$
and ${\bf 2}_{-1/2}$ respectively and are constant on $CP^2_F$.
$SU(3)$ singlets can now be constructed in the usual manner:
\begin{equation}
\hat D_R^\dagger {\cal D}\left( \Phi^\dagger \hat Q_L\right) +
\hat U_R^\dagger {\cal U} \left(\Phi_C^\dagger \hat Q_L\right)
+ \hat E_R^\dagger {\cal L} \left( \Phi^\dagger \hat L_L\right),
\label{Yukawa}
\end{equation}
where ${\cal D}$, ${\cal U}$ and ${\cal L}$ are arbitrary
complex $3\times 3$ matrices of Yukawa couplings.
The usual argument to derive the CKM matrix goes through without
modification  --- $U(3)$ rotations act on the left of the fermion
fields to diagonalise ${\cal U}$ and ${\cal L}$ and the CKM matrix
is in ${\cal D}$.

The point of this discussion is to show that all the usual arguments 
used to derive the CKM matrix work with this fuzzy construction
because they only require acting on the fermion generations from the left
by $U(3)$, and this does not interfere with the fact that we have
diagonalised $\hat\zeta^\dagger\hat\zeta$ by acting on $\hat\zeta$ 
from the right.  All the standard
arguments for the neutrino sector go through as well --- one can introduce
three Dirac and three Majorana neutrino masses as well as twelve complex
couplings \cite{SchechterValle}.
Of course the individual eigenstates here will not be zero-modes of the Dirac operator,
in general, and more work is necessary to determine whether or not
this suggestion for the origin of the generations is viable.

\section{Conclusions}

It has been argued that one generation of the standard model can be
obtained as a zero-mode of the Dirac operator on $CP^2\times CP^3$,
by introducing fundamental $U(2)\times U(3)$ gauge fields
and identifying the spin connection with the gauge connection.
The discussion circumvents Witten's no-go theorem for chiral fermions
in Kaluza-Klein theories precisely by introducing fundamental gauge
fields --- indeed fundamental gauge are essential as soon as
one considers $CP^2$ because $CP^2$ does not admit spinors
without them.
Three generations can be obtained by considering the $SU(3)$ isometry
group of $CP^2$ to be related to generation symmetry.  
Introducing concepts from
non-commuting geometry and making the complex projective spaces fuzzy
allows one to represent the zero-modes as finite matrices and
distorting the metric on $CP^2$ away from the $SU(3)$ symmetric metric
can lead to $3\times 3$ matrices whose eigenstates we identify with
the three generations.

The picture presented here borrows from many ideas that
are in the air at the moment, but it modifies them slightly
and puts them together in a rather different way than usual.
Kaluza-Klein theory uses the isometry group as the gauge
group, but here it is the holonomy group.  In Connes' non-commutative
version of the standard model two copies of space-time are introduced
to accommodate the two-component Higgs field while here three copies
of space-time are being introduced to accommodate three generations.
Also the three copies used here are being directly related
to a fuzzy space and I am not aware of any such interpretation
in the literature of the Connes-Lott model --- though the two copies used
there do look very like a fuzzy sphere and may have such an interpretation
(associating the generations with different copies of space-time
was suggested in Ref. \cite{A-HS}).
Another difference between the construction presented here and the
Connes-Lott model is that the gauge symmetries here are not automorphisms
of the matrix algebra.

Many questions remain to be investigated in this approach.
The holonomy group of $CP^2\times CP^3$ is $U(2)\times U(3)$,
which has two $U(1)$ factors, but only one linear
combination, dictated by (\ref{Ycharge}), has been used. 
This seems natural in view of the structure in equations
(\ref{square}) and (\ref{twothreestar}), but it raises the question
of the significance, if any, of the orthogonal combination
of $U(1)$ generators, which remains open.
The Higgs' fields and Yukawa couplings were 
introduced by hand here but it would clearly be 
preferable to have a geometrical interpretation --- it would be
satisfying if the Yukawa couplings could be incorporated into the
Dirac operator on the fuzzy spaces as in the Connes-Lott model.
The r\^ole of the $SU(4)$ isometry group of $CP^3$
has not been discussed here either --- it is tempting the think that it may
be related to the $4\times 4$ structure of equation (\ref{square}) 
in some, way but this remains to be investigated.
\vskip \baselineskip
{\bf Acknowledgments}:
when this talk was presented at Balfest in Vietri sul Mare, many people
asked stimulating questions and made useful suggestions.  In
particular I would like to thank Balachandran himself as well as
Giorgio Immirzi, Denjoe O'Connor, Pierre Ramond, 
Joe Schechter and Raphael Sorkin for useful comments.

\end{document}